# Epitaxial Graphene Intercalation: A Route to Graphene Modulation and Unique 2D Materials

Natalie Briggs[1,2,3], Zewdu M. Gebeyehu[1,4,5], Alexander Vera[1,2], Tian Zhao[6], Ke Wang[7], Ana De La Fuente Duran[1], Brian Bersch[1,2], Timothy Bowen[1], Kenneth L. Knappenberger, Jr.[6], Joshua A. Robinson[1-3,8]

[1]Department of Materials Science & Engineering, Pennsylvania State University, University Park, PA 16802
[2]Center for 2-Dimensional and Layered Materials, Pennsylvania State University, University Park, PA 16802
[3]2-Dimensional Crystal Consortium Materials Innovation Platform, Pennsylvania State University, University Park, PA 16802
[4]Catalan Institute of Nanoscience and Nanotechnology (ICN2), CSIC, The Barcelona Institute of Science and Technology (BIST), Campus UAB, Bellaterra, Spain, Barcelona, Spain
[5]Universitat Autònoma de Barcelona (UAB), Bellaterra, Spain
[6]Department of Chemistry, Pennsylvania State University, University Park, PA 16802
[7]Materials Characterization Laboratory, University Park, PA 16802
[8]Center for Atomically-Thin Multifunctional Coatings, Pennsylvania State University, University Park, PA 16802

## Abstract
Intercalation of atomic species through epitaxial graphene layers began only a few years following its initial report in 2004.[1] The impact of intercalation on the electronic properties of the graphene is well known; however, the intercalant itself can also exhibit intriguing properties not found in nature. This suggests that a shift in the focus of epitaxial graphene intercalation studies may lead to fruitful exploration of many new forms of traditionally 3D materials. In the following forward-looking review, we summarize the primary techniques used to achieve and characterize EG intercalation, and introduce a new, facile approach to readily achieve metal intercalation at the graphene/silicon carbide interface. We show that simple thermal evaporation-based methods can effectively replace complicated synthesis techniques to realize large-scale intercalation of non-refractory metals. We also show that these methods can be extended to the formation of compound materials based on intercalation. Two-dimensional (2D) silver (2D-Ag) and large-scale 2D gallium nitride (2D-$GaN_x$) are used to demonstrate these approaches.

## Introduction
Since its discovery >150 years ago, intercalation has enabled the creation of materials for applications ranging from catalysis and energy storage, to superconductivity and lubrication.[2–7] Importantly, intercalation also enables the decoupling of layers from bulk materials, as well as from native substrates.[8,9] For example, epitaxial graphene (EG) layers grown on silicon carbide (SiC) substrates may be physically decoupled from the substrate by intercalation to the EG/SiC interface.[9] EG is created when Si atoms sublime from the surface of SiC, leaving behind carbon atoms which reconstruct epitaxially with the SiC surface. This process enables the realization of large-area graphene layers, but also results in the formation of an electrically inactive, carbon-rich buffer layer which is partially covalently bound to the SiC surface.[9] This buffer layer scatters carriers in overlying graphene, thereby decreasing carrier mobility compared to exfoliated graphene.[10,11] However, intercalation of atomic species to the buffer layer/SiC interface both physically and electronically decouples the buffer layer from SiC, resulting in the formation of quasi-free standing epitaxial graphene (QFEG) layers with improved carrier transport.[12] This decoupling is possible with a wide range of intercalant species, including H, O, Ge, and Pd, among others.[9,13–20] While the precise mechanism of atomic intercalation to the EG/SiC interface is still under investigation, there is experimental evidence that the intercalant atoms diffuse through graphene defects and domain edges to reach to the EG/SiC interface.[9,15,21] In this review we discuss intercalation in EG/SiC systems, highlighting key techniques to realize intercalated EG/SiC materials. We emphasize the promise of intercalation beyond improved graphene properties, as a means of realizing unique, 2D intercalant layers and compounds such as 2D-Ag and 2D-$GaN_x$.

## Experimental Considerations for Intercalated Epitaxial Graphene

Experimental parameters play a significant role in achieving intercalation to the EG/SiC interface. Here we discuss key components of the EG intercalation process, including EG growth, surface treatments, precursor choice, and intercalation method.

## Epitaxial Graphene on Silicon Carbide: Growth and Characterization

While high-temperature graphitization of SiC was first demonstrated in the 1960s, the experimental realization of graphene in 2004 has led to a resurgence in the technique for mono- to few-layer graphene growth.[1,22,23] Heating SiC substrates to temperatures above 1200°C[24] under pressures ranging from ultra-high vacuum,[23] to atmospheric[24] results in the sublimation of silicon atoms from the SiC surface, where remaining carbon undergoes a $(6\sqrt{3} \times 6\sqrt{3})R30°$ reconstruction. This first carbon layer is commonly referred to as the buffer or zero layer, and consists of carbon atoms in a honeycomb structure, 1/3 of which are covalently bound to Si in the underlying SiC. Heating SiC substrates for increased time at elevated temperatures enables the formation of multiple graphene layers, where the continued sublimation of Si from the SiC leads to the creation of multiple graphene layers (Figure 1d).[25]

Identification and investigation of as-grown and intercalated EG layers on SiC is commonly performed through a variety of analytical techniques including low energy electron diffraction (LEED), as well as x-ray photoelectron (XPS) and angle-resolved photoemission spectroscopies (ARPES). When investigating intercalation, low energy electron diffraction may be used to monitor the EG/SiC system as it progresses from its initial state, containing a $(6\sqrt{3} \times 6\sqrt{3})R30°$ reconstructed (buffer) layer bonded to SiC (Figure 1d), to its final intercalated state, where the buffer layer is decoupled from the SiC and transformed to QFEG (Figure 1e).[26] Because the buffer layer exhibits a markedly different electron pattern from QFEG, the two structures may be easily distinguished (Figure 1a,b). Apart from LEED, changes in the C 1s and Si 2p core electron regions acquired from XPS can also indicate intercalation (Figure 1c). The C1s region of pristine EG/SiC exhibits features related to the buffer layer, bulk SiC, as well as any present QFEG. Given sufficient energy resolution, the C 1s buffer layer component (Figure 1c) may be fitted with S1 and S2 peaks, where S1 corresponds to carbon atoms in the reconstructed buffer layer bound to one Si atom in SiC and 3 $sp^2$ C atoms within the buffer layer. S2 corresponds to the remaining $sp^2$ C atoms in the reconstructed layer.[26] Intercalation of atoms to the EG/SiC interface eliminates the buffer layer component from the C 1s region, due to disruption of the buffer layer bonding to SiC. Additionally, a shift of approximately 0.7-2.0eV in bulk SiC C 1s and Si 2p peak positions is often observed in the XPS spectrum as a result of intercalation.[9,15,18,27–35] This shift indicates a change in the charge transfer between SiC and EG caused by the presence of an intercalant layer.

Changes in the doping of graphene layers can also signify intercalation. This can be investigated through ARPES, where the graphene Dirac point can be monitored relative to the Fermi level. Changes in the Dirac point following exposure of EG to intercalant species can indicate intercalation to the EG/SiC interface and decoupling of the buffer layer (Figure 1f).[9] Similarly, restoration of the Dirac point to its original energy following additional sample annealing can indicate deintercalation of atoms from the EG/SiC interface and reformation of the buffer layer (Figure 1f,v).[9] While practically absent from intercalation studies to-date, Auger electron spectroscopy and cross-sectional scanning transmission electron microscopy are also valuable techniques to directly investigate the surface and interfacial chemistry of intercalated EG/SiC systems in addition to the structure of the intercalant layer.[36–38] As the field progresses, these techniques should be considered key components for investigation and identification of intercalated materials.

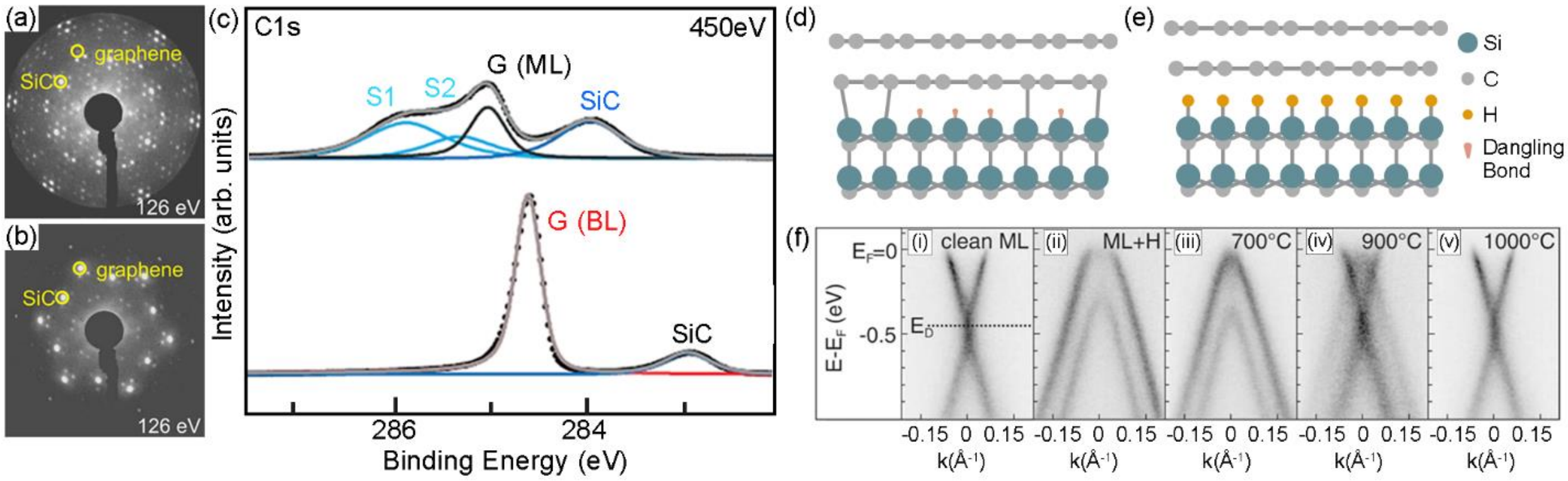

Figure 1: LEED patterns for buffer layer EG before (a) and after (b) hydrogen intercalation, where reconstruction spots are suppressed upon intercalation of hydrogen atoms. (c) C 1s core level spectra of monolayer graphene (buffer + 1 layer) before (top) and after (bottom) hydrogen intercalation, where hydrogen intercalation decouples the buffer layer, creating QFEG, as depicted in (d) and (e). The C 1s SiC peak shifts to lower binding energies upon intercalation and the buffer layer S1 and S2 components are no longer present. (f) ARPES measurements show the monolayer graphene (d) Dirac cone before (f, i) and after (f, ii) hydrogen intercalation. Deintercalation of hydrogen is observed as the sample is annealed, and the Dirac point returns to its initial energy (f, iii-v). Images are reproduced with permission from reference 9.

## Defect-Mediated Intercalation

Epitaxial graphene can readily accommodate intercalation.[14,15,18,19,39,40] However, intercalation through pristine, multilayer EG is limited by defects native to the grown graphene layers. Thus, the low defect density afforded by high-temperature EG synthesis techniques can limit the lateral scale of intercalation. Successful intercalation is attributed to the formation of cracks and defects in graphene layers caused by initial deposition of intercalant species on the sample surface.[15] However, reproducible, large-area intercalation can be achieved through the use of EG layers which are highly defective to begin with, for example, as a result of exposure to plasma treatments. Exposing EG layers to $O_2$/He plasmas for instance can result in the formation of carbon vacancies, where vacancy edge atoms may also become passivated by the plasma constituents. Passivated species such as C-O-C and C=O are believed to aid intercalation by binding to intercalant atoms more strongly than pristine graphene layers.[37] Thus, these passivated species can effectively draw intercalants to the graphene surface where intercalation can take place. Plasma-treated EG layers are used to realize large-area Ga, In, and Sn,[37,38] as well as Ag and $GaN_x$ demonstrated later in this review.

**Intercalation Approaches**

A wide library of EG/SiC intercalants have been demonstrated over the last decade (Table 1), where intercalants enable p-type, n-type, as well as charge neutral graphene. Ge and Au have been shown to enable both p and n type graphene doping individually, which is attributed to different thicknesses of the elemental layers at the EG/SiC interface. Most EG intercalations (Table 1) utilize an initial deposition step, where metal atoms such as Co, Pt, Fe, and Au are deposited onto the EG surface via thermal or E-beam evaporation, sputtering, or molecular beam/Knudsen cell deposition.[19,27,34,40] Following this step, EG/SiC substrates are annealed at temperatures typically ≥600°C under ultra-high vacuum (UHV), to induce intercalation.[19,27,32,34,40] This synthesis approach is well suited to investigation of intercalated structures, due to the relative ease of characterizing samples *in situ* through LEED, ARPES, XPS, or a combination of all three techniques. Investigation of intercalated materials through scanning tunneling microscopy and spectroscopy is also common, and can reveal ordered intercalant structures underneath the graphene layers.[16,28,41–44]

Elemental intercalation may also be achieved using gas-phase precursors such as $H_2$, $O_2$, air, and $NH_3$.[9,17,31,32,36] In a similar fashion to metals deposited on EG, gas phase atomic species are hypothesized to adsorb onto the graphene surface and diffuse through graphene layers. In contrast to typical UHV metal intercalation, intercalation of elements from gaseous precursors is typically performed near atmospheric pressure.[9,17,32,45] While less studied, alternative sources such as plasmas, metal-organics, molten baths, and molecular precursors may also be used to achieve intercalation.[13,36,46,47] Such sources have been successfully used to intercalate H, Ga, Ca and F. Overall, these studies demonstrate that a wide range of precursors and pressures may be used to achieve elemental intercalation.

Ultra-high vacuum metal intercalation is the most widely utilized EG intercalation technique, but recent studies have shown that metal intercalation can also be achieved through simple, non-UHV methods. For example, metal intercalation can be achieved by heating a solid precursor and an EG/SiC substrate together in a tube furnace or sealed ampoule.[37,47,48] This simultaneous heating approach is effective for intercalating a wide range of non-refractory metals with melting temperatures near or below ~900°C, which can be easily reached in a simple tube furnace. Heating at or above the melting temperature of the metallic precursor often enables a vapor pressure great enough for intercalation. However, to intercalate high-melting temperature metals, the use of compound precursors may be necessary. In such cases, decomposition chemistry, relative constituent vapor pressure, and flow or vapor transport characteristics must also be considered to achieve successful intercalation.

**Table 1: EG/SiC Intercalation Conditions (Temperature, Pressure) and Resulting Graphene Carrier Type**

| Intercalant | Intercalation Pressure | Intercalation Temperature (°C) | Deintercalation Temperature (°C) | Majority Carrier Type | $E_D$ relative to $E_f$ ($E_f$=0) (meV) |
|---|---|---|---|---|---|
| H | 600 Torr[12]<br>atmospheric[9] | 600 – 1200[9,12] | | p | 100[9] |
| Li | UHV[30,49] | 290 – 330[15]<br>350[49]<br>360[30] | 500[49] | n | -650 – -900[15]<br>-1000[49]<br>-1400[30] |
| N | | 500[31] | | | |
| O | Atmospheric[17,32,45,50] | 250[17,50]<br>600[32,45]<br>750[50] | | p | |
| F | | 200[51]<br>800[13] | 1200[13] | p[51]<br>neutral[13] | 790[51]<br>0[13] |
| Na | UHV[52] | 180[52] | | n | |
| Si | UHV | 750[33]<br>800[53] | 1000[33,53] | n | -260[33]<br>-300[53] |
| Ca | | 350[47] | | | |
| Mn | UHV | 600[54] | 1200[54] | n | -300[54] |
| Fe | UHV | 600[34] | | n | -250[34] |
| Co | UHV | 650 – 800[16,40] | | | |
| Cu | UHV | 600[39]<br>700[20] | 800[39] | n | 900[39]<br>850[20] |
| Ga | 50 Torr | 550, 675[36] | | | |
| Ge | UHV | 720[14] | 800[14] | p, n | |
| Pd | UHV | >700[18] | 900[18] | neutral | |
| Sn | UHV | 850[35] | 1050[35] | | |
| Eu | UHV | 800[55]<br>120 – 300[56] | 1050[56] | n | |
| Yb | UHV[57] | 500[57] | | n | 1500[57] |
| Pt | UHV | 900[27] | | n | -150[27] |
| Au | UHV[19,41,58] | 800 – 1000[19,41,58] | | p[19], n[19,58] | 100[19]<br>-232[58]<br>-850[19] |
| Pb | UHV | 675[28] | | p | 100[28] |

**Intercalation via Thermal Vaporization: Creating 2D Silver and $GaN_x$**

Silver is a promising material for plasmon-based technologies such as biosensors and waveguides.[59–64] However, implementation of Ag films into plasmonic devices is limited by the need for encapsulation layers that can protect Ag films from oxidation. Intercalation of Ag atoms through EG layers is one promising method to realize Ag films with an inherent, overlying graphene capping layer.[65,66] While intercalation of Ag through graphene on native metal substrates has been performed with mixed success,[67–69] we show that EG/SiC may be reliably intercalated with Ag to realize encapsulated, 2D-Ag films for plasmonic applications.

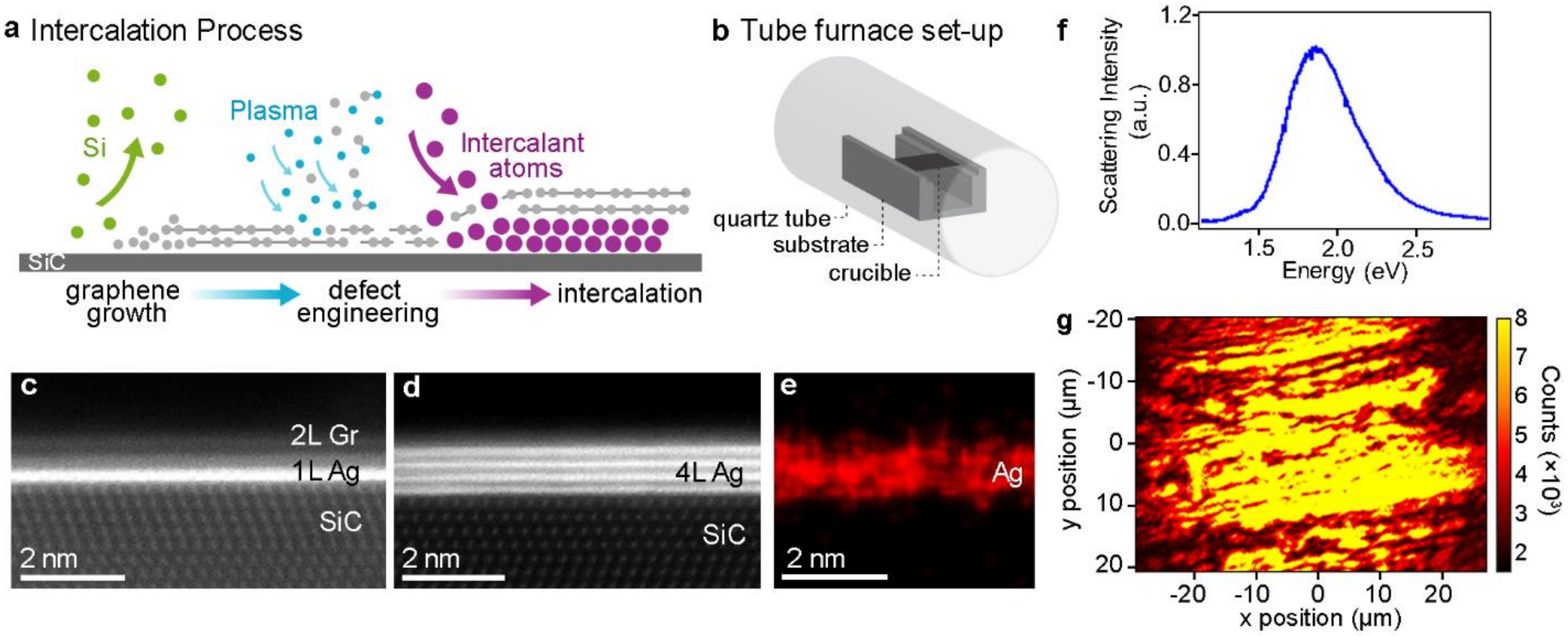

Figure 2: (a) Intercalation schematic illustrating graphene growth, plasma treatment, and metal intercalation (b) Diagram of experimental setup depicting a quartz tube containing an alumina crucible and an EG/SiC substrate placed facing downward, into the crucible. (c) Cross-sectional STEM showing 1 and (d) 4 layers of Ag between EG and SiC. (e) Corresponding Ag EDS. (f) Dark field scattering spectrum showing a peak at 1.87 eV, reflecting the plasmonic character of the Ag film. (g) Map of SHG peak across the EG/Ag/SiC surface showing the presence of Ag layers across 10s of microns.

2D-Ag films are synthesized using the process detailed in Figure 2a, where EG is first synthesized on SiC.[70] Following this synthesis, defects in the graphene layers are generated by exposing graphene to an $O_2$/He plasma treatment.[37,48] Graphene/SiC substrates are then placed face down over metallic Ag powder and heated to 900-950°C, under 300 Torr for 20 minutes (Figure 2b). A constant flow rate of 50 sccm Ar is maintained throughout the synthesis period. Cross-sectional scanning transmission electron microscopy (STEM) of intercalated Ag samples shows 1 and 4 Ag layers located between graphene and the SiC substrate (Figure 2c,d). These interfacial layers are confirmed to consist of Ag atoms through energy dispersive x-ray spectroscopy mapping (EDS) (Figure 2e). Dark-field microscopy and second harmonic generation (SHG) imaging measurements are performed to investigate the plasmonic properties of 2D-Ag. The normalized dark-field scattering spectrum in Figure 2f shows a peak at 1.87 eV which reveals that the intercalated Ag film exhibits a plasmon resonance in the visible region. The same measurements are performed on a 40 nm, bare Ag film directly deposited on SiC via E-beam evaporation. Unlike the intercalated Ag, the deposited film does not show dark field spectral peak, possibly due to oxidation of the Ag film during transportation of the sample and measurement in ambient. Thus, intercalation of Ag atoms to the EG/SiC interface is a promising approach for protection of metal films from oxidation.

Second harmonic generation imaging microscopy (Figure 2g) confirms that 2D-Ag created at the EG/SiC interface supports non-linear optical phenomena with uniform responses over large areas. Decreased intensity observed near the edges of the image is attributed to the point-spread function of the optical setup utilized for the measurement. However, the stripe-like pattern is the result of 2D-Ag discontinuities that result from large (>2 nm) terrace steps in the SiC substrate. This demonstrates that even with uniform defect formation in EG prior to intercalation, the SiC morphology can play a dominant role in achieving uniform elemental intercalation. The nonlinear optical enhancement observed via SHG measurements further supports the notion that intercalated Ag layers exist in a metallic, plasmon-supporting state and are protected from oxidation by overlying graphene.[71,72] The SHG measurements also enable extraction of the second order non-linear susceptibility ($\chi^2$) of 2D-Ag. In the case of 2D-Ag formed via intercalation, $\chi^2 \approx 1.7 \times 10^{-9}$ m/V, which is 250× larger than that of Ag nanoparticle colloids.[73] The large increase in $\chi^2$ is attributed to a combination of plasmon-resonant excitation and the non-symmetric bonding that occurs in 2D metals intercalated at the interface of EG and SiC[37] that enables large second-order non-linear responses. Therefore, the creation of 2D-Ag constitutes a major advancement in the use of 2D Ag films in practical photonic applications. Beyond the realization of 2D Ag films, these studies show that this simple thermal vaporization-based approach readily enables the realization of new, 2D materials, and constitutes a relatively unutilized method for further 2D materials discovery, beyond exfoliation and chemical vapor deposition techniques.

Thermal vaporization and reaction can also be applied to realize 2D compound materials at the EG/SiC interface. The creation of 2D $GaN_x$ initially reported by Al Balushi *et al.*, utilized Ga intercalation from trimethyl gallium, followed by reaction with $NH_3$.[36] However, 2D $GaN_x$ can readily be realized through simple evaporation of a metallic Ga source followed by annealing in $NH_3$, *in situ* or *ex situ*. The process is identical to that outlined in Figure 2a and b, but intercalation and reaction is performed at a temperature of 700°C. Following synthesis, samples are characterized via cross-sectional STEM, EDS, and electron energy loss spectroscopy (EELS) (Figure 3a-e). Figure 3a,b shows two cross-sectional STEM regions, where the first consists of the characteristic Ga bilayer observed in previous 2D $GaN_x$ studies. The second region includes thicker $GaN_x$ located between graphene and SiC. This is also consistent with previous 2D-$GaN_x$ studies, which show a range of GaN thicknesses at the EG/SiC interface. Both 2D-$GaN_x$ and thicker $GaN_x$ show N and Ga signatures from EDS and EELS measurements (Figure 3c,d,e). These results demonstrate an alternative approach to realizing compound 2D materials between graphene and SiC which does not require complex or logistically difficult metal organic sources. Thus, this thermal evaporation-based method is a rapid, accessible route for the synthesis of both 2D metals and 2D compounds at the EG/SiC interface.

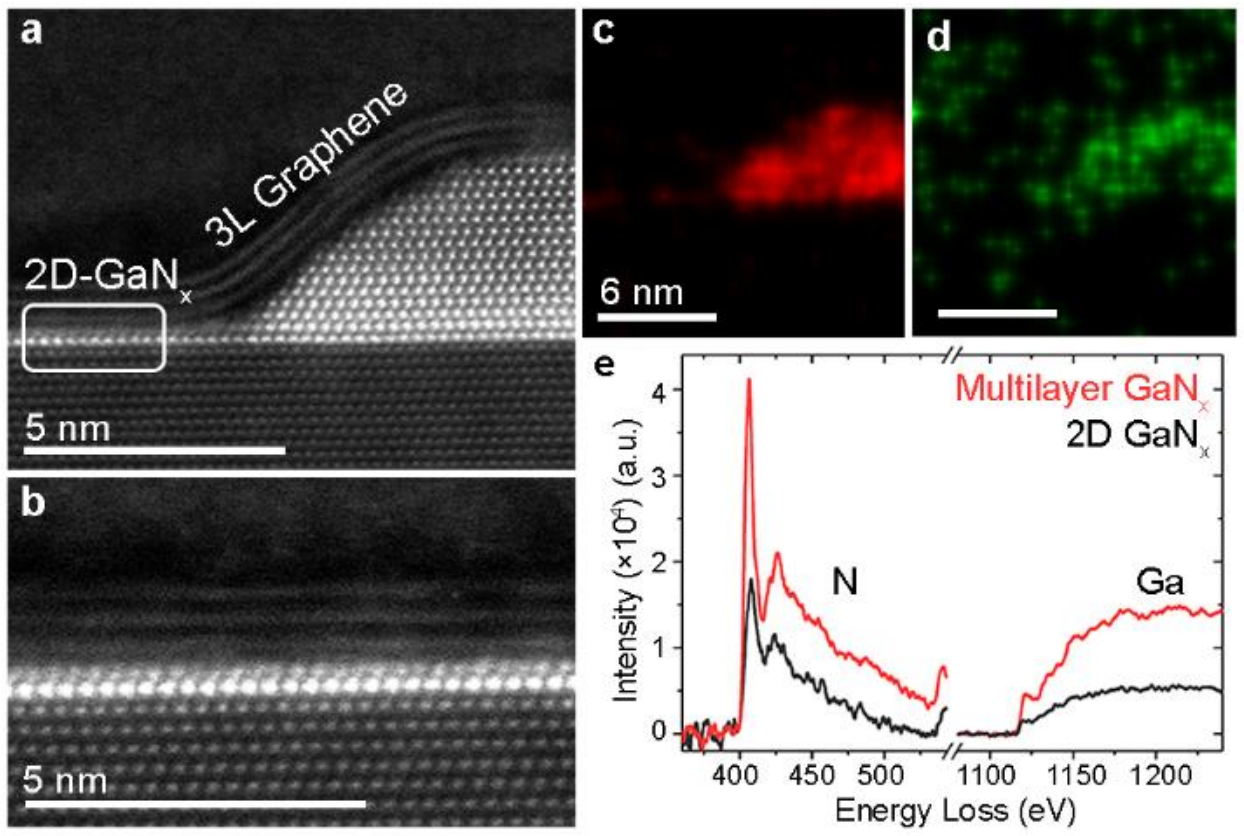

Figure 3: (a, b) Cross-sectional STEM showing 2D and multilayer $GaN_x$ between EG and SiC. (c,d) EDS maps of Ga and N for the multilayer region shown in (a). (e) EELS spectra for $GaN_x$ in (a,b) showing N and Ga signatures from both regions.

## Conclusion and Outlook

Over the last decade, EG/SiC intercalation has been demonstrated under UHV and atmospheric conditions using a variety of precursor materials. Recent studies emphasize the importance of graphene defects in achieving large-area intercalation, and demonstrate a simplified approach to realizing EG/metal/SiC structures. This approach utilizes thermal evaporation of metal sources at 300 Torr to realize large-area, plasmonically-active metal layers which are protected from environmental effects. Looking forward, intercalation and reaction at the EG/SiC interface may be adopted to realize a wide-array of intercalated systems, where the intercalant may be investigated beyond the context of its impact on overlying graphene. Furthermore, 2D layers and bulk substrates beyond graphene and SiC may be investigated as host systems to enable new, 2D layers. This approach can enable the realization of 2D forms of materials that do not exist in nature, and hold promise in technologies ranging from biosensors to superconductors.

## Acknowledgements

This work was supported in part by the 2D Crystal Consortium National Science Foundation (NSF) Materials Innovation Platform under cooperative agreement DMR-1539916, the Semiconductor Research Corporation Intel/Global Research Collaboration Fellowship, task 2741.001, and by a grant from the Air Force Office of Scientific Research, grant number FA-9550-18-1-0347.

## References

1. Berger, C. *et al.* Ultrathin Epitaxial Graphite: 2D Electron Gas Properties and a Route toward Graphene-based Nanoelectronics. *J. Phys. Chem. B* **108**, 19912–19916 (2004).
2. Boersma, M. A. M. Catalytic properties of alkali metal-graphite intercalation compounds. *Catal. Rev.* **10**, 243–280 (1974).
3. Brodie, B. C. XIII . On the atomic weight of graphite. *Philos. Trans. R. Soc. London* **149**, 249–259 (1859).
4. Schafhaeutl, C. Ueber die Verbindungen des Kohlenstoffes mit Silicium. *J. Chem. Inf. Model.* **53**, 1689–1699 (1840).
5. Dresselhaus, M. S. & Dresselhaus, G. Intercalation compounds of graphite. *Adv. Phys.* **30**, 139–326 (1981).
6. Ebert, L. B. Catalysis by graphite intercalation compounds. *J. Mol. Catal.* **15**, 275–296 (1982).
7. Enoki, T., Suzuki, M. & Endo, M. *Graphite intercalation compounds and applications*. (Oxford University Press, 2003).
8. Zhang, Z. & Lerner, M. M. Preparation, Characterization, and Exfoliation of Graphite Perfluorooctanesulfonate. *Chem. Mater.* **8**, 257–263 (1996).
9. Riedl, C., Coletti, C., Iwasaki, T., Zakharov, A. A. & Starke, U. Quasi-Free-Standing Epitaxial Graphene on SiC Obtained by Hydrogen Intercalation. *Phys. Rev. Lett.* **103**, 1–4 (2009).
10. Bolotin, K. I. *et al.* Ultrahigh electron mobility in suspended graphene. *Solid State Commun.* **146**, 351–355 (2008).
11. Jobst, J. *et al.* Quantum oscillations and quantum Hall effect in epitaxial graphene. *Phys. Rev. B* **81**, 195434 (2010).
12. Robinson, J. A. *et al.* Epitaxial Graphene Transistors: Enhancing Performance via Hydrogen. *Nano Lett.* **11**, 3875–3880 (2011).
13. Wong, S. L. *et al.* Quasi-Free-Standing Epitaxial Graphene on SiC (0001) by Fluorine Intercalation from a Molecular Source. *ACS Nano* **5**, 7662–7668 (2011).
14. Emtsev, K. V, Zakharov, A. A., Coletti, C., Forti, S. & Starke, U. Ambipolar doping in quasifree epitaxial graphene on SiC (0001) controlled by Ge intercalation. *Phys. Rev. B* **84**, 1–6 (2011).

15. Virojanadara, C., Watcharinyanon, S., Zakharov, A. A. & Johansson, L. I. Epitaxial graphene on 6H-SiC and Li intercalation. *Phys. Rev* **82**, 1–6 (2010).
16. de Lima, L. H., Landers, R. & de Siervo, A. Patterning Quasi-Periodic Co 2D-Clusters underneath Graphene on SiC(0001). *Chem. Mater.* **26**, 4172–4177 (2014).
17. Oida, S. *et al.* Decoupling graphene from SiC(0001) via oxidation. *Phys. Rev. B - Condens. Matter Mater. Phys.* **82**, 1–4 (2010).
18. Yagyu, K., Takahashi, K., Tochihara, H., Tomokage, H. & Suzuki, T. Neutralization of an epitaxial graphene grown on a SiC (0001) by means of palladium intercalation. *Appl. Phys. Lett.* **110**, 1–6 (2017).
19. Gierz, I. *et al.* Electronic decoupling of an epitaxial graphene monolayer by gold intercalation. *Phys. Rev. B - Condens. Matter Mater. Phys.* **81**, (2010).
20. Forti, S. *et al.* Mini-Dirac cones in the band structure of a copper intercalated epitaxial graphene superlattice. *2D Mater.* **3**, 035003 (2016).
21. Markevich, A. *et al.* First-principles study of hydrogen and fluorine intercalation into graphene-SiC(0001) interface. *Phys. Rev. B - Condens. Matter Mater. Phys.* **86**, 1–9 (2012).
22. Novoselov, K. S. *et al.* Electric field effect in atomically thin carbon films. *Science* **306**, 666–9 (2004).
23. Badami, D. V. Graphitization of α-Silicon Carbide. *Nature* **193**, 569–570 (1962).
24. Emtsev, K. V. *et al.* Towards wafer-size graphene layers by atmospheric pressure graphitization of silicon carbide. *Nat. Mater.* **8**, 203–207 (2009).
25. Riedl, C., Coletti, C. & Starke, U. Structural and electronic properties of epitaxial Graphene on SiC(0001): A review of growth, characterization, transfer doping and hydrogen intercalation. *J. Phys. D. Appl. Phys.* **43**, (2010).
26. Riedl, C. Epitaxial Graphene on Silicon Carbide Surfaces: Growth, Characterization, Doping and Hydrogen Intercalation. *Ph.D. Friedrich-Alexander-Universität Erlangen-nürnb.* 1–173 (2010).
27. Xia, C., Johansson, L. I., Niu, Y., Zakharov, A. A. & Janze, E. High thermal stability quasi-free-standing bilayer graphene formed on 4H-SiC(0001) via platinum intercalation. *Carbon N. Y.* **79**, 631–635 (2014).
28. Yurtsever, A., Onoda, J., Iimori, T., Niki, K. & Miyamachi, T. Effects of Pb Intercalation on the Structural and Electronic Properties of Epitaxial Graphene on SiC. *Small* **12**, 3956–3966 (2016).
29. Sieber, N. *et al.* Synchrotron x-ray photoelectron spectroscopy study of hydrogen-terminated 6H-SiC{0001} surfaces. *Phys. Rev. B* **67**, 205304 (2003).
30. Bisti, F. *et al.* Electronic and geometric structure of graphene / SiC (0001) decoupled by lithium intercalation. **245411**, 1–7 (2015).
31. Wang, Z. *et al.* Simultaneous N-intercalation and N-doping of epitaxial graphene on 6H-SiC (0001) through thermal reactions with ammonia. *Nano Res.* **6**, 399–408 (2013).
32. Oliveira, M. H. *et al.* Formation of high-quality quasi-free-standing bilayer graphene on SiC (0001) by oxygen intercalation upon annealing in air. *Carbon N. Y.* **52**, 83–89 (2012).
33. Silly, M. G. *et al.* Electronic and structural properties heterostructures engineered by silicon intercalation. *Carbon N. Y.* **76**, 27039 (2014).
34. Sung, S. J. *et al.* Spin-induced band modifications of graphene through intercalation of magnetic iron atoms. *Nanoscale* **6**, 3824–3829 (2014).
35. Kim, H., Dugerjav, O. & Lkhagvasuren, A. Charge neutrality of quasi-free-standing monolayer graphene induced by the intercalated Sn layer. *J. Phys. D. Appl. Phys.* **49**, 1–7 (2016).
36. Al Balushi, Z. Y. *et al.* Two-dimensional gallium nitride realized via graphene encapsulation. *Nat. Mater.* **15**, 1166–1171 (2016).
37. Briggs, N. *et al.* Confinement Heteroepitaxy: Realizing Atomically Thin, Half-van der Waals Materials. *http://arxiv.org/abs/1905.09962* (2019).
38. Bersch, B. *et al.* An Air-Stable and Atomically Thin Graphene/Gallium Superconducting Heterostructure. *https://arxiv.org/abs/1905.09938* (2019).

39. Yagyu, K. *et al.* Fabrication of a single layer graphene by copper intercalation on a SiC(0001) surface. *Appl. Phys. Lett.* **104**, (2014).
40. Zhang, Y., Zhang, H., Cai, Y., Song, J. & He, P. The investigation of cobalt intercalation underneath epitaxial graphene on 6H-SiC(0 0 0 1). *Nanotechnology* **28**, 075701 (2017).
41. Premlal, B. *et al.* Surface intercalation of gold underneath a graphene monolayer on SiC(0001) studied by scanning tunneling microscopy and spectroscopy. *Appl. Phys. Lett.* **94**, 1–4 (2009).
42. Cranney, M. *et al.* Superlattice of resonators on monolayer graphene created by intercalated gold nanoclusters. *EPL (Europhysics Lett.* **91**, 66004 (2010).
43. Nair, M. N. *et al.* High van Hove singularity extension and Fermi velocity increase in epitaxial graphene functionalized by intercalated gold clusters. *Phys. Rev. B - Condens. Matter Mater. Phys.* **85**, 1–6 (2012).
44. Zhang, Y., Zhang, H., Cai, Y., Song, J. & He, P. The investigation of cobalt intercalation underneath epitaxial graphene on 6H-SiC(0 0 0 1). *Nanotechnology* **28**, 075701 (2017).
45. Kowalski, G., Tokarczyk, M., Ciepielewski, P. & Baranowski, J. M. SiC (0001) New X-ray insight into oxygen intercalation in epitaxial graphene grown on 4 H -SiC (0001). **105301**, (2017).
46. Virojanadara, C., Zakharov, A. A., Yakimova, R. & Johansson, L. I. Buffer layer free large area bi-layer graphene on SiC(0 0 0 1). *Surf. Sci.* **604**, L4–L7 (2010).
47. Li, K. *et al.* Superconductivity in Ca-intercalated epitaxial graphene on silicon carbide. *Appl. Phys. Lett.* **103**, 1–3 (2013).
48. Bersch, B. *et al.* An Air-Stable and Atomically Thin Graphene/Gallium Superconducting Heterostructure. *submitted* (2019).
49. Virojanadara, C., Zakharov, A. A., Watcharinyanon, S., Yakimova, R. & Johansson, L. I. A low-energy electron microscopy and x-ray photo-emission electron microscopy study of Li intercalated into graphene on SiC(0001). *New J. Phys.* **12**, 125015 (2010).
50. Ostler, M. *et al.* Decoupling the Graphene Buffer Layer from SiC(0001) via Interface Oxidation. *Mater. Sci. Forum* **717–720**, 649–652 (2012).
51. Walter, A. L. *et al.* Highly p-doped epitaxial graphene obtained by fluorine intercalation. *Appl. Phys. Lett.* **98**, 1–3 (2011).
52. Sandin, A. *et al.* Multiple coexisting intercalation structures of sodium in epitaxial graphene-SiC interfaces. *Phys. Rev. B* **125410**, 1–5 (2012).
53. Xia, C. *et al.* Si intercalation/deintercalation of graphene on 6H-SiC(0001). *Phys. Rev. B* **85**, 045418 (2012).
54. Gao, T. *et al.* Atomic-Scale Morphology and Electronic Structure of Manganese Atomic Layers Underneath Epitaxial Graphene on SiC (0001). *ACS Nano* **6**, 6562–6568 (2012).
55. Anderson, N. A., Hupalo, M., Keavney, D., Tringides, M. C. & Vaknin, D. Intercalated europium metal in epitaxial graphene on SiC. *Phys. Rev. Mater.* **1**, 054005 (2017).
56. Sung, S. *et al.* Observation of variable hybridized-band gaps in Eu-intercalated graphene. *Nanotechnology* **28**, 205201 (2017).
57. Watcharinyanon, S., Johansson, L. I., Xia, C. & Virojanadara, C. Ytterbium oxide formation at the graphene–SiC interface studied by photoemission. *J. Vac. Sci. Technol. A Vacuum, Surfaces, Film.* **31**, 020606 (2013).
58. Nair, M. N. *et al.* High van Hove singularity extension and Fermi velocity increase in epitaxial graphene functionalized by intercalated gold clusters. *Phys. Rev. B* **85**, 245421 (2012).
59. Johnson, P. B. & Christy, R. W. Optical Constants of the Noble Metals. *Phys. Rev. B* **6**, 4370–4379 (1972).
60. Tsuda, Y., Omoto, H., Tanaka, K. & Ohsaki, H. The underlayer effects on the electrical resistivity of Ag thin film. *Thin Solid Films* **502**, 223–227 (2006).
61. Anker, J. N. *et al.* Biosensing with plasmonic nanosensors. *Nat. Mater.* **7**, 442–453 (2008).
62. Simon, H. J., Mitchell, D. E. & Watson, J. G. Optical Second-Harmonic Generation with Surface Plasmons in Silver Films. *Phys. Rev. Lett.* **33**, 1531–1534 (1974).

63. Quail & Simon. Second-harmonic generation from silver and aluminum films in total internal reflection. *Phys. Rev. B. Condens. Matter* **31**, 4900–4905 (1985).
64. Oulton, R. F., Sorger, V. J., Genov, D. A., Pile, D. F. P. & Zhang, X. A hybrid plasmonic waveguide for subwavelength confinement and long-range propagation. *Nat. Photonics* **2**, 496–500 (2008).
65. Losurdo, M. *et al.* Graphene as an Electron Shuttle for Silver Deoxidation: Removing a Key Barrier to Plasmonics and Metamaterials for SERS in the Visible. *Adv. Funct. Mater.* **24**, 1864–1878 (2014).
66. Hong, H. Y., Ha, J. S., Lee, S.-S. & Park, J. H. Effective Propagation of Surface Plasmon Polaritons on Graphene-Protected Single-Crystalline Silver Films. *ACS Appl. Mater. Interfaces* **9**, 5014–5022 (2017).
67. Tontegode, A. Y. & Rut'kov, E. V. Intercalation by atoms of a two-dimensional graphite film on a metal. *Physics-Uspekhi* **36**, 1053–1067 (1993).
68. Starodubov, A. G., Medvetskii, M. A., Shikin, A. M. & Adamchuk, V. K. Intercalation of silver atoms under a graphite monolayer on Ni(111). *Phys. Solid State* **46**, 1340–1348 (2004).
69. Farías, D., Shikin, A. M., Rieder, K.-H. & Dedkov, Y. S. Synthesis of a weakly bonded graphite monolayer on Ni(111) by intercalation of silver. *J. Phys. Condens. Matter* **11**, 8453–8458 (1999).
70. Subramanian, S. *et al.* Properties of synthetic epitaxial graphene/molybdenum disulfide lateral heterostructures. *Carbon N. Y.* **125**, 551–556 (2017).
71. Hong, H. Y., Ha, J. S., Lee, S.-S. & Park, J. H. Effective Propagation of Surface Plasmon Polaritons on Graphene-Protected Single-Crystalline Silver Films. *ACS Appl. Mater. Interfaces* **9**, 5014–5022 (2017).
72. Kravets, V. G. *et al.* Graphene-protected copper and silver plasmonics. *Sci. Rep.* **4**, 5517 (2015).
73. Kityk, I. V *et al.* Nonlinear optical properties of Au nanoparticles on indium–tin oxide substrate. *Nanotechnology* **16**, 1687–1692 (2005).